\documentclass[12pt,a4paper]{article}

  \usepackage{hyperref}
  \usepackage{a4wide}
  \usepackage{latexsym}
  \usepackage{epsf}
  \usepackage{amssymb}
  \usepackage{graphicx}
  \usepackage{cite}
  \usepackage{slashed,epsfig}
  \usepackage{bbm}
  \usepackage{color}
  \usepackage{amsmath,amssymb,amsthm}
  \usepackage{psfrag}

\renewcommand{\d}{\textrm{d}}
\newcommand{\Real}{\textrm{I\!R}}

\renewcommand{\d}{\textrm{d}}

\newcommand{\SU}{\mathop{\rm SU}}
\newcommand{\SO}{\mathop{\rm SO}}

\newcommand{\GL}{\mathop{\rm GL}}

\newcommand{\eq}[1]{\begin{equation}#1\end{equation}}
\newcommand{\spl}[1]{\begin{split}#1\end{split}}
\newcommand{\al}[1]{\begin{align}#1\end{align}}
\newcommand{\subeq}[1]{\begin{subequations}#1\end{subequations}}

\begin{document}

\begin{flushright}
\small
KUL-TF-10/03\\
UUITP-07/10
\date \\
\normalsize
\end{flushright}

\begin{center}

\vspace{.7cm}

{\LARGE {\bf Universal de Sitter solutions at tree-level}} \\

\vspace{2cm}

{\large Ulf H. Danielsson$^a$, Paul Koerber$^b$ and  Thomas Van
Riet$^a$} \footnote{ulf.danielsson \emph{at} physics.uu.se, koerber
\emph{at}
itf.fys.kuleuven.be, thomas.vanriet \emph{at} fysast.uu.se } \\
\vspace{1cm}

$^a$ {\small\slshape Institutionen f\"{o}r Fysik och Astronomi,\\
Box 803, SE-751 08 Uppsala, Sweden}
\\\vspace{.3cm}
$^b$ {\small\slshape Instituut voor Theoretische Fysica, K.U.\ Leuven, \\Celestijnenlaan 200D, B-3001 Leuven, Belgium} \\

\vspace{2cm}

{\bf Abstract} \end{center} {\small Type IIA string theory
compactified on $\SU(3)$-structure manifolds with orientifolds
allows for classical de Sitter solutions in four dimensions. In this
paper we investigate these solutions from a ten-dimensional point of
view. In particular, we demonstrate that there exists an attractive
class of de Sitter solutions, whose geometry, fluxes and source
terms can be entirely written in terms of the universal forms
that are defined on all $\SU(3)$-structure manifolds. These are the forms $J$ and
$\Omega$, defining the $\SU(3)$-structure itself, and the torsion
classes. The existence of such universal de Sitter solutions is
governed by easy-to-verify conditions on the $\SU(3)$-structure,
rendering the problem of finding dS solutions purely geometrical. We
point out that the known (unstable) solution coming from the
compactification on $\SU(2)\times\SU(2)$ is of this kind.}

\newpage

\pagestyle{plain} 

\section{Introduction}
Flux compactifications of IIA string theory with O6-planes allow for
supersymmetric (susy) anti-de Sitter (AdS) vacua with all moduli
stabilised at tree-level in both the string coupling $g_S$ and
$\alpha'$ \cite{DeWolfe:2005uu}. A natural, cosmologically more
relevant, extension of this would be to achieve moduli stabilisation
at tree-level in a de Sitter (dS) background instead. We have to
take care, though, to circumvent the no-go theorems\footnote{The
first of these no-go theorems for de Sitter solutions was
established in \cite{Maldacena:2000mw}, requiring negative-tension
sources, like O-planes, to circumvent it. It has been later extended
in \cite{Hertzberg:2007wc} by investigating the dependence of the 4d
scalar potential on the dilaton and volume modulus. This has then
been used to construct de Sitter no-go theorems tailored to various
models \cite{Silverstein:2007ac, Haque:2008jz, Caviezel:2008tf,
Flauger:2008ad, Danielsson:2009ff, Wrase:2010ew}. See also
\cite{Covi:2008cn} for a different approach.} that make the
construction of a de Sitter solution a difficult task. Some of the
ingredients required for circumventing de Sitter no-go theorems are
already present in the setup of \cite{DeWolfe:2005uu}, such as
non-zero Romans mass ($F_0$-flux) and O6-planes. A further necessary
ingredient is negative curvature of the internal space
\cite{Silverstein:2007ac, Haque:2008jz}. Negative curvature is not
present in the original model of \cite{DeWolfe:2005uu} since it
assumed an internal Calabi--Yau space. Fortunately, the
generalisation of the supersymmetric AdS solution of
\cite{DeWolfe:2005uu} turns out to be a compactification on
an $\SU(3)$-structure manifold \cite{Lust:2004ig,tomasiellocosets,Koerber:2008rx,
Caviezel:2008ik, Cassani:2009ck}, which is generically curved. As a
consequence, de Sitter solutions in this generalised setup where
found in three different papers \cite{Danielsson:2009ff,
Flauger:2008ad,Caviezel:2008tf}. We briefly discuss these three
solutions as they play an important role in the forthcoming.

The solutions of \cite{Caviezel:2008tf} and \cite{Flauger:2008ad}
are very similar in nature. They are both established numerically by
minimising the scalar potential in four dimensions. The internal
space is a group manifold with orbifold group $\mathbb{Z}_2\times
\mathbb{Z}_2$ and four intersecting smeared O6-planes. The solution
found in \cite{Caviezel:2008tf} uses the group manifold
$\SU(2)\times\SU(2)$ while \cite{Flauger:2008ad} contains two other
examples, and the one that is most relevant to us is based on the
solvable group s1.2. Although these examples \emph{are the first
examples of classical 4d de Sitter solutions ever constructed}, they
are perturbatively unstable and therefore not useful as true
(metastable) vacua.

Ref.~\cite{Danielsson:2009ff} constructed a class of 4d de Sitter
solutions, directly in 10 dimensions, with the same ansatz used for
the susy AdS solution, albeit with different flux quanta, such that
the solution breaks supersymmetry and has positive energy. The
condition for having a solution to the ten-dimensional equations of
motion translates in a simple condition on the torsion classes
\cite{Danielsson:2009ff}. However \cite{Danielsson:2009ff} did not
provide an explicit manifold with torsion classes that satisfies
this condition in the case of a de Sitter-space. The aim of this
paper is to amend this, which requires us to generalise the universal de
Sitter solutions of \cite{Danielsson:2009ff}. This situation is
similar to the general susy AdS solutions, where first the conditions
were constructed in a general way \cite{Lust:2004ig}, and later
more geometries (next to the solutions of \cite{cveticnk1}),
satisfying all the constraints on the torsion
classes, where found \cite{tomasiellocosets,Koerber:2008rx}.

This ten-dimensional approach based on universal forms has some advantages with respect to
the four-dimensional approach. First, it is technically easier than
minimising a multiple-variable function. Second, it is conceptually
attractive since it describes a solution irrespective of the details
of the geometry. It therefore describes a \emph{whole class of
solutions}\footnote{Of course, issues like perturbative stability
can only be addressed once a specific geometry is used.}. Third, one
is assured that the solution solves the ten-dimensional equations of
motion and, on the same footing, issues like flux and charge
quantisation require a lift to ten dimensions anyway.

The generalisation of the universal de Sitter solutions of
\cite{Danielsson:2009ff} we describe in this paper, has a  more
general $\SU(3)$-structure. The susy AdS solutions of
\cite{Lust:2004ig,tomasiellocosets,Koerber:2008rx,Caviezel:2008ik}  have vacuum
expectation values for the geometric moduli for which only two
torsion classes ($W_1, W_2$) are non-zero. In this paper we
generalise by allowing the third torsion class ($W_3$) to be
non-zero as well. The key result we have found is a geometry that
explicitly satisfies the geometrical constraints for the universal
de Sitter solutions, thereby proving that universal de Sitter
solutions exist. Remarkably, we rediscover the known solution on the
$\SU(2)\times\SU(2)$ group manifold, thereby providing the lift of
the solution to ten dimensions. But the main message of this paper
is not about the details of that solution, rather, the proof of
principle that the set of universal de Sitter solutions is non-zero.
This probably indicates that there must be many more geometries that
allow the universal de Sitter solutions, and we hope to report on a
more thorough investigation of this soon \cite{progress}.

However not all de Sitter solutions should be universal since, as we
explain below, the solvmanifold solution \cite{Flauger:2008ad} is
not of this kind. Nonetheless, there is a simple argument why one
expects the universal solutions to be more likely. The Ricci tensor
of $\SU(3)$-structure manifolds is entirely expressed in terms of
the universal forms \cite{bedulli-2007-4, Ali:2006gd}. From the
Einstein equation this implies that the energy-momentum tensor should be
written in terms of contractions of these forms. The most natural
way to achieve this is to assume that all the fluxes and sources are
given in terms of these forms. Indeed, not only the known susy AdS
solutions are universal, recently also classes of non-susy AdS
solutions were found using a universal ansatz
\cite{Lust:2008zd,Cassani:2009ck,dimitriosextrasol,Danielsson:2009ff,Koerber:2010rn}\footnote{This was
furthermore used for constructing Lifshitz solutions in type II
supergravity \cite{Li:2009pf,Blaback:2010pp}. In this case solutions
were found in type IIA* supergravity and not in IIA supergravity
\cite{Blaback:2010pp}.}. In \cite{Koerber:2010rn} it was even shown
that sourceless solutions exist that can achieve moduli
stabilisation.

This paper is organised as follows. In section \ref{preliminaries}
we establish our conventions for SU(3)-structure and the
type IIA supergravity equations of motion. In section \ref{universal}
we present the universal de Sitter solutions, whereas in section
\ref{groupmanifolds} we investigate the conditions for universal de
Sitter solutions on group manifolds with $\SU(3)$-structure and
establish that the group $\SU(2)\times \SU(2)$ fulfills all the
constraints. The solution on $\SU(2)\times \SU(2)$ is discussed in
some detail in section \ref{SU2XSU2}. Finally in section
\ref{discussion} we end with a discussion.

\section{Preliminaries}\label{preliminaries}
\subsection{$\SU(3)$-structure manifolds}

A six-dimensional $\SU(3)$-structure manifold can be characterised
by a globally defined real two-form $J$ and a complex decomposable
three-form $\Omega=\Omega_R+i\Omega_I$, satisfying a compatibility and a
normalisation condition \eq{ \Omega \wedge J=0 \, , \qquad
\Omega\wedge\Omega^*=(4i/3) \, J\wedge J\wedge J = 8i \label{normcond} \,
\text{vol}_6 \, . } From the real part of the three-form we can build an almost complex
structure for which $J$ is of type $(1,1)$ and $\Omega$ is of type
$(3,0)$. It is given by
\begin{equation}\label{complexstructure}
I^l_{\,\,k}= c \, \varepsilon^{m_1m_2\ldots
m_5l}(\Omega_R)_{km_1m_2}(\Omega_R)_{m_3m_4m_5}\,,
\end{equation}
where $\varepsilon$ is the Levi-Civita symbol, and the real scalar $c$ is
such that $I$ is properly normalised: $I^2= -\mathbbm{1}$.
The metric then follows via
\begin{equation}\label{metricfromI}
g_{mn}=-I^l_{\,\,m}J_{ln}\,.
\end{equation}

The torsion classes $W_1,\ldots, W_5$ correspond to the expansion of
the exterior derivatives of $J$ and $\Omega$ in terms of
SU(3)-representations \cite{chiossal}. In this paper we will put
$W_4=W_5=0$ and take $W_1,W_2$ real (in our conventions), leading to
a so-called {\em half-flat manifold}. We will motivate this
truncation in section \ref{groupSU3}. We find then
\subeq{\label{tcl}
\begin{align}
&\d J = \frac{3}{2}W_1\Omega_R + W_3\,,\\
&\d\Omega_R = 0 \, , \\
&\d\Omega_I = W_1 J\wedge J + W_2\wedge J\, ,
\end{align}
}
where $W_1$ is a real scalar, $W_2$ a
real primitive $(1,1)$-form and $W_3$ a real primitive $(1,2)+(2,1)$-form. This means
that
\subeq{\al{
\hspace{1cm} & W_2\wedge J\wedge J=0\, , && W_3 \wedge J = 0 \, , \hspace{1cm} \\
& W_2\wedge \Omega=0 \, ,  &&  W_3 \wedge \Omega = 0 \, . }}
Furthermore, we find that under the Hodge star, defined from the
metric \eqref{metricfromI}, \eq{\label{hodgeprop}
\star_6\Omega=-i\Omega \, , \qquad \star_6 J=\tfrac{1}{2}\, J\wedge
J  \, , \qquad \star_6 W_2=-J\wedge W_2 . }

We will make use of the fact that the Ricci tensor can be expressed
in terms of the torsion classes \cite{bedulli-2007-4} (see also \cite{Ali:2006gd}).
Let us first observe that any real symmetric two-tensor $T_{ij}$ splits as
follows in representations of SU(3) \eq{\label{symdecomp} T_{ij} =
\frac{s(T_{ij})}{6} g_{ij} + T_{ij}^+ + T_{ij}^- \, . } Here
$s(T_{ij})$ is the trace, an SU(3)-invariant, and $T^+_{ij}$
and $T^-_{ij}$ transform as $\bf{8}$ and $\bf{6}+\bf{\bar{6}}$
respectively. The latter are traceless and have respectively index
structure (1,1) and (2,0)+(0,2) \subeq{\al{
& T_{ij}^+ g^{ij} = 0 \, , \qquad I^i{}_k T^+_{ij} I^j{}_l = T^+_{kl} \, , \\
& T_{ij}^- g^{ij} = 0 \, , \qquad I^i{}_k T^-_{ij} I^j{}_l =
-T^-_{kl} \, . }} Furthermore to $T_{ij}^+$ and $T_{ij}^-$ a
primitive real (1,1)-form and a complex primitive (2,1)-form can be
respectively associated \subeq{\label{symdecomp2}\al{
\label{symtwoformpart} & P_2(T_{ij}) = \frac{1}{2} J^k{}_i T^+_{kj} \, \d x^{i} \wedge \d x^j   \, , \\
\label{symthreeformpart} & P_3(T_{ij}) = \frac{1}{2} T^-_{il} \Omega^l{}_{jk} \,
\d x^{i} \wedge \d x^j \wedge \d x^{k} \, . }}

Using this decomposition it was shown in \cite{bedulli-2007-4} that
the Ricci tensor can be expressed as follows in terms of the torsion
classes \subeq{\al{
s(R_{ij}) & = \frac{15}{2} (W_1)^2 - \frac{1}{2} (W_2)^2 -\frac{1}{2} (W_3)^2 \, , \\
P_2(R_{ij}) & = - \frac{1}{4} \star (W_2 \wedge W_2)  - \frac{1}{2} \star_6 d \star_6 \left(W_3 - \frac{1}{2} W_1 \Omega_R\right) \, , \label{ricci2} \\
P_3(R_{ij}) & = 2 \, W_1 W_3|_{(2,1)} + 2 \, \d W_2|_{(2,1)} - \frac{1}{4}
Q_1(W_3,W_3) \, , \label{ricci3}
}}
with \eq{ Q_1(W_3,W_3) = \left(\Omega^{ijk}
\iota_j \iota_i W_3 \wedge \iota_k W_3\right)_{(2,1)} \, , }
and where in the right-hand side of eqs.~\eqref{ricci2}-\eqref{ricci3}
the projection onto the primitive part is understood.

Finally it will be convenient to extract the norms of $W_2$ and $W_3$ as
follows \subeq{\al{
W_2 & = w_2 \hat{W}_2 \, , \qquad w_2 = \sqrt{(W_2)^2} \, , \\
W_3 & = w_3 \hat{W}_3 \, , \qquad w_3 = \sqrt{(W_3)^2} \, , }} where
we introduced $(W_2)^2 = \frac{1}{2} W_{2ij} W_2^{ij}$ and likewise $(W_3)^2 = \frac{1}{3!} W_{3ijk}
W_3^{ijk}$.

\subsection{IIA supergravity with calibrated sources}

We have to solve the equations of motion of type IIA supergravity,
for which we will use the string-frame equations listed in appendix B of
\cite{Koerber:2007hd}, and put $2 \, \kappa_{10}^2=1$. These equations contain source terms and are
valid when the sources are calibrated (which must be the case if the sources are embedded supersymmetrically).

The dilaton $\Phi$ and the warp factor $A$ are taken to be constant (actually we take $A=0$)
and the fluxes $F_n$ ($n=0,2,4,6,8,10$ in the democratic formalism \cite{Bergshoeff:2001pv})
are decomposed as follows in an ``electric'' and a ``magnetic'' part
\eq{ F = \hat{F} +
\text{vol}_4 \wedge \tilde{F}\, ,}
where $\hat{F}$ and $\tilde{F}$ have only internal indices.
The self-duality constraint of the democratic formalism relates the electric
and the magnetic flux so that it suffices to calculate the magnetic part in the following.
In the presence of calibrated D6/O6 sources $j$, we can write the
non-trivial equations of motion as follows {\allowdisplaybreaks
\subeq{\label{eomsugra}\al{
& \d \hat{F}_2 + H \hat{F}_0 = - j,  && (\text{Bianchi }\hat{F}_2) \\
& \d \star_6 \hat{F}_4 - H \wedge \star_6 \hat{F}_6=0, && (\text{eom }\hat{F}_4)  \\
& \d(e^{-2\Phi} \star_6 H) - (\star_6 \hat{F}_2) \hat{F}_0 - (\star_6 \hat{F}_4) \wedge \hat{F}_2 - (\star_6 \hat{F}_6)\wedge \hat{F}_4 =0, &&  (\text{eom }H) \\
& 2(R_{4}+R_6) - H^2 - e^{\Phi} \star_6 (\Omega_I \wedge j) = 0,  &&  (\text{eom $\Phi$}) \\
& R_{4} + e^{2\Phi} \sum_n (\tilde{F}_{(n)}^2) + e^{\Phi} \star_6 (\Omega_I \wedge j)=0, && (\text{external Einstein}) \\
& - \tfrac{1}{2} H^2 + \tfrac{1}{4}e^{2\Phi} \sum_n (5-n)
\hat{F}_{(n)}^2 + \tfrac{3}{4} e^{\Phi} \star_6 (\Omega_I \wedge j)
=0,
&& (\text{Tr Einstein/eom $\Phi$})\\
& R_{ij} -\tfrac{1}{2} H_i \cdot H_j - \tfrac{1}{4}e^{2\Phi} \sum_n \left(\hat{F}_{(n)i} \cdot \hat{F}_{(n)j} - \tilde{F}_{(n)i} \cdot \tilde{F}_{(n)j}\right) \nonumber \\
& + \tfrac{1}{4}e^{\Phi} \left\{ - g_{ij} \star_6 (\Omega_I \wedge j
) + 2 \star_6 \left[(g_{k(i} \d x^k \wedge \iota_{j)}
\Omega_I)\wedge j\right]\right\}=0, && (\text{Einstein/eom $\Phi$})
\label{eomeinstein}}}}
where we defined $\phi_{i} \cdot \phi_{j} = \iota_i \phi \cdot \iota_j \phi=
\frac{1}{(l-1)!} \phi_{ii_2 \ldots i_l} \phi_{j}{}^{i_2\ldots i_l}$
for an $l$-form $\phi$. As compared to the equations of motions of
\cite{Koerber:2007hd}, we found it convenient to take linear
combinations of the Einstein equation, its trace and the dilaton
equation of motion.

\section{Universal de Sitter solutions in general}\label{universal}

The basic idea of universal de Sitter solutions is to find solutions
independent of the details of the compactification manifold. This
can be done by writing down an ansatz in terms of form-fields that
are defined on all $\SU(3)$-structure manifolds. Then the equations
of motion translate into simple conditions on these form-fields.
Once these simple conditions are satisfied for a given manifold, one
has an explicit solution in ten dimensions. In
ref.~\cite{Danielsson:2009ff} such an ansatz was
presented,\footnote{Compared to \cite{Danielsson:2009ff} we have
absorbed some convenient dilaton factors in the definition of the
flux parameters, ensuring that all dependence on the dilaton drops
out of the equations of motion. This implies that the dilaton is a
free parameter in the solutions we will present.}
\subeq{\label{fluxansatz1}\al{
\hspace{2cm} e^{\Phi}\hat{F}_0 & = a_1 \, ,  & e^{\Phi} \hat{F}_2  &= a_2 J + a_3 \hat{W}_2 \, , \hspace{2cm} \\
e^{\Phi}\hat{F}_4 & = a_4\, J \wedge J\,, & e^{\Phi }\hat{F}_6 &= a_5 \text{vol}_6 \, , \\
H & = a_6 \, \Omega_R  \, ,  & e^{\Phi} j   &= j_1 \Omega_R  \, , }}
which also covers the supersymmetric AdS
solutions\footnote{The susy AdS solution of \cite{Lust:2004ig} is given by
$a_1=e^{\Phi} m\,,a_2=-W_1/4\,, a_3=-w_2\,,a_4=3\, e^{\Phi}m/10\,,
a_5=9\,W_1/4\,, a_6=2m e^{\Phi}/5$ and $j_1=-2e^{2\Phi}m^2/5
+ 3(W_1^2-2w_2^2)/8$. The ansatz can be extended to also allow for a term
proportional to $J \wedge \hat{W}_2$ in $\hat{F}_4$, which was used in \cite{Koerber:2010rn} to construct
non-susy AdS solutions.}. To find a
solution different from the susy AdS solution one has to impose
\cite{Danielsson:2009ff}
\begin{equation}
\d \hat{W}_2=c_1\Omega_R\,,\qquad \hat{W}_2\wedge \hat{W}_2=
c_2J\wedge J + d_2 \hat{W}_2\wedge J\,,
\end{equation}
where $c_1, c_2, d_2$ are real proportionality coefficients.
The coefficients $c_1$ and $c_2$ are fixed by internal consistency of the
$\SU(3)$-structure equations,
\begin{equation}
c_1=-\frac{w_2}{4}\,,\qquad c_2=-\frac{1}{6}\,,
\end{equation}
while $d_2$ is specific to the geometry. The equations of
motion then fix all geometrical quantities in terms of the flux parameters
$a_i$ and the charge $j_1$
\begin{equation}\label{geometricalvalues}
W_1=W_1(a_i,j_1)\,,\qquad w_2=w_2(a_i,j_1)\,,\qquad d_2=d_2(a_i,
j_1)\,.
\end{equation}
Furthermore, the parameters $a_i$ and $j_1$ are related amongst
themselves by the equations of motion, such that there is a slice in
the space formed by the variables $(a_i, j_1)$ that solves the
equations of motion. We have explicitly demonstrated in
\cite{Danielsson:2009ff} that there exist ranges of values for the
parameters $a_i$ and $j_1$ (on the parameter slice that solves the
equations) such that the resulting cosmological constant is
positive. However, up to now, we have not found a geometry that can
reach these values for the geometrical quantities
(\ref{geometricalvalues}).

\subsection{The ansatz}

Because we have not yet found a suitable geometry that allows for the universal
de Sitter solution (\ref{fluxansatz1}) we propose a
more general universal ansatz, which considers a half-flat manifold
and includes $W_3$ as an expansion form. In a forthcoming paper \cite{progress} we will
study this ansatz in general. It turns out, however, that one obtains a
very interesting simplified case if one puts
\begin{equation}
W_2=0\, ,
\end{equation}
which we will consider in this paper. The ansatz for the fluxes is then
\subeq{\label{fluxansatz2}\al{
\hspace{3cm} e^{\Phi}\hat{F}_0  & = f_1 \, , & e^{\Phi}\hat{F}_2 &= f_2 J \, , \hspace{3cm}\\
 e^{\Phi}\hat{F}_4 & = f_3 J \wedge J \, , & e^{\Phi}\hat{F}_6  &= f_4 \text{vol}_6 \, , \\
H & = f_5 \Omega_R + f_6 \hat{W}_3 \, , & e^{\Phi}j  &= j_1 \Omega_R
+ j_2 \hat{W}_3 \, . }}
For dS-solutions the source $j$ will correspond to O6-planes. When writing such an ansatz,
one has to take into account the parity under the orientifold involutions. For O6-planes we
have that the fluxes $H, \hat{F}_2$ and $\hat{F}_6$ are odd whereas $\hat{F}_0$ and $\hat{F}_4$
are even. Furthermore, if the O6-planes are supersymmetrically embedded and thus preserve
the SU(3)-structure we have that $J$, $\Omega_R$ and $\hat{W}_3$
are odd, therefore our ansatz is justified from that point of view.

Upon plugging the ansatz in the equations of motion one
immediately finds certain constraints on the remaining torsion classes
\subeq{\label{tclprop}\al{
& \d \star_6 \hat{W}_3 = c_1 J \wedge J  \, , \\
& (\hat{W}_{3\, i} \cdot \hat{W}_{3\, j})^+ = 0\, , \label{tclprop2}
}}
where we find $c_1=w_3/3!$ from computing $\d \star_6 \hat{W}_3 \wedge J$.
 With the use of these constraints, the equations of motion \eqref{eomsugra} lead to a set of
algebraic relations for the constants $f_i$ and $j_1, j_2$
{\allowdisplaybreaks \subeq{\label{eomeinsteinexp}\al{ & \left( \,\tfrac{3}{2}\, f_2 W_1 +
f_1 f_5 + j_1 \right) \Omega_R +
\left( f_2 w_3 +  f_1 f_6 + j_2 \right) \hat{W_3} =0 \label{bianchiF2}\, , \\
& \left( 3 f_3 W_1  -f_4 f_5 \right) \Omega_R +\left( 2 f_3 w_3 -
f_4 f_6 \right) \hat{W}_3 = 0
\label{eomF_4} \, , \\
& \left( f_5 W_1 +  \tfrac{1}{6} f_6 w_3 - \tfrac{1}{2} f_1 f_2 - 2 f_2 f_3 - f_3 f_4 \right) J \wedge J =0\label{eomH} \\
& R_4 = - \tfrac{15}{2} (W_1)^2  + \tfrac{1}{2}\left[(w_3)^2+(f_6)^2 \right] + 2 f_5^2 + 2 j_1 \,, \label{dilaton eom}\\
& R_4 + (f_1)^2 + 3(f_2)^2  + 12 (f_3)^2 + (f_4)^2 + 4 j_1 = 0\,, \label{externalEinstein}\\
& -2 (f_5)^2 - \tfrac{1}{2} (f_6)^2 + \tfrac{1}{4} \left[ 5 (f_1)^2 + 9 (f_2)^2 +12 (f_3)^2 - (f_4)^2 \right] +3 j_1 = 0\,,\label{trace Einstein/dilatoneom} \\
& 2 (W_1 w_3  -  j_2 - 2f_5 f_6) \hat{W}_3|_{(2,1)} -
\tfrac{1}{4}\, \left[w_3^2 \, Q_1(\hat{W}_3,\hat{W}_3)+f_6^2\, Q_2(\hat{W}_3,\hat{W}_3)\right] = 0
\label{Einsteinthree-formpart}\,,}}} where
\begin{equation}
Q_2(\hat{W}_3,\hat{W}_3) = \left.\left(\frac{1}{2} \hat{W}_{3\,imn}
\hat{W}_3{}^{pmn} \Omega_{pjk} \, \d x^i \wedge \d x^j \wedge \d x^k\right)\right|_{(2,1)}\,.
\end{equation}
We note that using the decomposition \eqref{symdecomp}-\eqref{symdecomp2}, eq.~\eqref{eomeinstein}
would normally have a two-form and a three-form part. The two-form part is satisfied using the constraint \eqref{tclprop2},
while the three-form part leads to \eqref{Einsteinthree-formpart} above.

We see that the three-form piece
of the internal Einstein equation leads to further constraints.
We will put
\begin{equation}
Q_1(\hat{W}_3,\hat{W}_3) = c_2 Q_2 (\hat{W}_3,\hat{W}_3)=
c_3(\hat{W}_3)_{2,1} \,.
\end{equation}
The equations \eqref{eomeinsteinexp} will then lead to relations between the coefficients $f_i, j_1, j_2$
and the parameters $w_3,c_2,c_3$ characterising the geometry.

\subsection{The solutions}
\label{solana}

We will now study the solutions to the equations \eqref{eomeinsteinexp}, in particular
those with $R_4 > 0$, which are the dS solutions. We will demonstrate later on that with
the O-plane configuration of section \ref{groupSU3} one obtains
\begin{equation}
Q_1 = Q_2 \Rightarrow c_2=1 \, , \qquad c_3=\frac{8}{\sqrt{3}}\, .
\end{equation}
For these choices one can show that $f_3$ and $f_4$ have to be zero
in order to find a de Sitter solution\footnote{De Sitter solutions
to eqs.~\eqref{eomeinsteinexp} also exist when
$c_3\neq \frac{8}{\sqrt{3}}$ and then one can take $f_3, f_4$
non-zero. The geometry should then, however, fall outside the class
studied in section \ref{groupmanifolds} and we have not found such
an example.}. Once $f_3=f_4=0$ the general solution to the equations
forms a two-dimensional set in the parameter space. The equations
are however invariant under an overall rescaling \eq{ f_i
\rightarrow \lambda f_i \, , \quad j_i \rightarrow \lambda^2 j_i \,
, \quad R_4 \rightarrow \lambda^2 R_4 \, , \quad W_1 \rightarrow
\lambda W_1\, , \quad w_3\rightarrow\lambda w_3 \, . }
We will choose to factor out this overall scale by dividing
each quantity by appropriate factors of $f_1$ (which is equivalent to considering
solutions with $f_1=1$). We then end up with a one-dimensional set of solutions that include
non-susy AdS solutions, non-susy Minkowski solutions \emph{and de
Sitter solutions}.

\begin{figure}
\centering
\psfrag{lambda}{$\scriptstyle \Lambda/(f_1)^2$}
\psfrag{f2}{$\scriptstyle f_2/f_1$}
\includegraphics[scale=0.6]{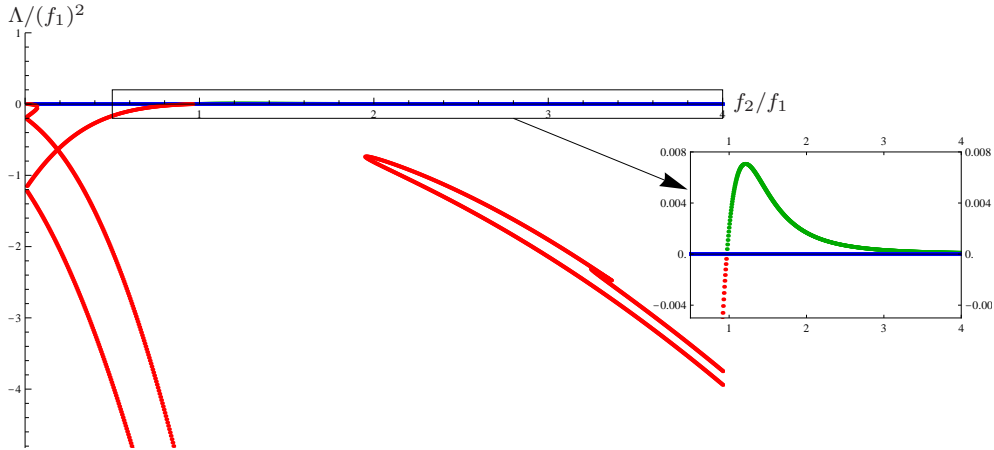}
\caption{$\Lambda/(f_1)^2$ as a function of $f_2/f_1$, for
$c_3=8/\sqrt{3}$ and $f_3=f_4=0$. The area close to the horizontal axis is shown in
more detail such that the dS solutions (in green) are visible. AdS solutions are
shown in red and Minkowski solutions in blue.}
\label{plotlambda}
\end{figure}

This is made clear in figure \ref{plotlambda}, where we plot the
value of the cosmological constant (vertical axes) against $f_2/f_1$
(horizontal axis). This plot contains all kinds of solutions, but to
make the de Sitter solutions visible, in an inset we zoomed in on an area very close
to the horizontal axis.
Most essential in the figure is the interval with de Sitter
solutions bounded from below by a Minkowski point at
$f_2/f_1 = 0.965$ and this forms our key result.
For large values of $f_2/f_1$ the line of de Sitter
solutions asymptotes to a line of Minkowski solutions, which are
remarkably simple solutions taking the form
\begin{equation}
f_5/f_1=\frac{1}{4}\,,\qquad f_6/f_1=\frac{\sqrt{3}}{2}\,,\qquad
W_1=\frac{f_2}{2}\,,\qquad w_3/W_1=3\sqrt{3}\,.\label{Minkowski}
\end{equation}
This phenomenon, where de Sitter solutions interpolate between
Minkowski solutions in parameter space, was first observed in
\cite{deCarlos:2009qm}, which obtained similar results from a
four-dimensional point of view. Here we gain the extra insight
of a ten-dimensional interpretation. On one side the bounding
four-dimensional Minkowski solutions are the remarkably
simple solutions (\ref{Minkowski}) with fluxes along the universal
forms. On the other side we find a Minkowski solution where the curve
of figure \ref{plotlambda} crosses the x-axis (as displayed in the inset).
For the de Sitter solutions itself the explicit forms of
the parameters $f_i$ are not that insightful, so apart from the plots
we do not present them explicitly.

Most abundant are the non-supersymmetric AdS branches in solution
space. It seems a generic property of tree-level flux
compactifications (allowing dS solutions) that the AdS solutions far
outnumber the dS solutions.

\begin{figure}
\psfrag{w3}{$\scriptstyle w_3/W_1$}
\psfrag{f2}{$\scriptstyle f_2/f_1$}
\psfrag{p}[tl]{\textcolor{blue}{$\scriptstyle (0.965,4.553)$}}
\centering
\includegraphics[scale=0.6]{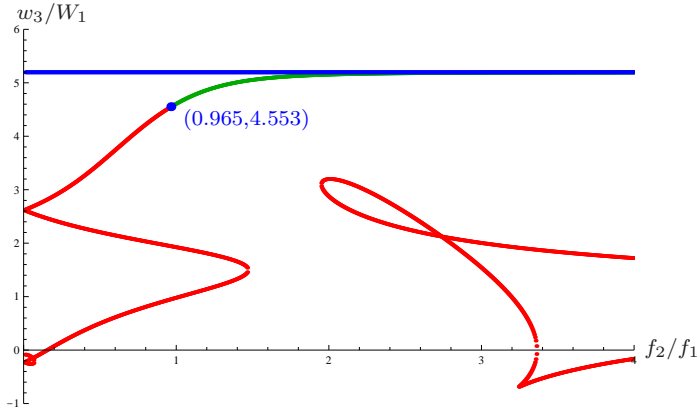}
\caption{$w_3/W_1$ as a function of $f_2/f_1$, for
$c_3=8/\sqrt{3}$. dS solutions are green, AdS solutions red and Minkowski solutions blue.}
\label{plotw3}
\end{figure}

Figure \ref{plotw3} displays the ratio $w_3/W_1$ on the vertical
axis and $f_2/f_1$ on the horizontal axis.
This plot reveals that the ratio $w_3/W_1$ is bounded from above for all solutions and
that the saturation value $w_3/W_1=3\sqrt{3}$ corresponds to the line of
Minkowski solutions \eqref{Minkowski}. Furthermore we read off from the plot that
in other to obtain dS solutions one must find a geometry such that
\eq{
4.553<w_3/W_1< 3 \sqrt{3} \, .
}

\section{Universal de Sitter solutions on group manifolds}\label{groupmanifolds}

\subsection{Group manifolds, orientifolds and $\SU(3)$-structures}\label{groupSU3}

Group manifolds form an interesting class of candidates for the internal manifold.
If the six-dimensional group manifold $G$ is not already compact we assume that there
exists a discrete subgroup $\Gamma
\subset G$, without fixed points, such that $G/\Gamma$ is a compact
manifold. This $\Gamma$ does not always exist and we refer to
\cite{Grana:2006kf} (and references therein) for a discussion on this. On $G$ we have six,
globally-defined, left-invariant one-forms (a.k.a.\ Maurer--Cartan
forms), $e^i$, with $i=1,\ldots,6$. These forms obey the
Maurer--Cartan relations
\begin{equation}
\d e^i=\tfrac{1}{2} f^i{}_{jk}e^j \wedge e^k \,,
\end{equation}
with $f$ the structure constants of the Lie algebra $\mathfrak{g}$
associated to $G$. We will restrict ourselves to the supergravity degrees of
freedom that are expanded along these left-invariant forms, which are named the
left-invariant degrees of freedom.

A compactification of a supergravity theory on such spaces leads to
a lower-dimensional supergravity theory with the same amount of
supersymmetries, if one restricts to the left-invariant degrees of
freedom. Therefore, if we desire de Sitter solutions in an
$\mathcal{N}=1$ supergravity theory we need at least three
intersecting O6-planes (implying a fourth). Naively extending the Tseytlin rules for
brane intersections in flat space to the case of a group manifold,
leads to the following intersection of four O6-planes (we present
only the internal directions)
\begin{center}
  \begin{tabular}{|c|c|c|c|c|c|}
    \hline
    \rule[1em]{0pt}{0pt} $e^1$ & $e^2$  & $e^3$  & $e^4$ & $e^5$  & $e^6$  \\
    \hline
    \hline
    \rule[1em]{0pt}{0pt} $\bigotimes$ & $\bigotimes$  & $\bigotimes$ & -- & -- & -- \\\hline
    \rule[1em]{0pt}{0pt} $\bigotimes$ & -- & -- & $\bigotimes$  & $\bigotimes$ & -- \\\hline
    \rule[1em]{0pt}{0pt} -- & $\bigotimes$ & -- & -- & $\bigotimes$  & $\bigotimes$ \\\hline
    \rule[1em]{0pt}{0pt} -- & -- & $\bigotimes$ & $\bigotimes$ & -- & $\bigotimes$ \\\hline
   \end{tabular}
\end{center}
where each entry denotes a left-invariant direction. This
intersection is unique up to relabeling of the O6-planes and
relabeling the Maurer--Cartan forms. From this we find that the
smeared orientifold source is given by
\begin{equation}
j_6 =j_Ae^{456} + j_Be^{236} + j_Ce^{134} + j_De^{125}\,,
\end{equation}
with the corresponding involutions
\eq{\label{oriinv}\spl{
&A  :\qquad (e^4, e^5, e^6) \rightarrow\,- (e^4, e^5, e^6)\,, \\
&B  :\qquad (e^2, e^3, e^6) \rightarrow\,- (e^2, e^3, e^6)\,, \\
&C  :\qquad (e^1, e^3, e^4) \rightarrow\,- (e^1, e^3, e^4)\,, \\
&D  :\qquad (e^1, e^2, e^5) \rightarrow\,- (e^1, e^2, e^5)\,.
}}
Note that
\begin{equation}
A.B.C=D\,.
\end{equation}
This shows that three O6-involutions, in this setting, imply the fourth.
Alternatively, one can look at this as one orientifold involution
(say $A$) together with the orbifold group
$\mathbb{Z}_2\times\mathbb{Z}_2$ generated by $AB$ and $BC$.
Therefore our compactification space is
\begin{equation}
\frac{G}{\Gamma\times\mathbb{Z}_2\times\mathbb{Z}_2}\,.
\end{equation}
When the orbifold singularities are blown up, we generate new moduli
commonly denoted as the twisted sector. We do not discuss this any
further, but a thorough analysis of moduli stabilisation should also
include these modes.

There are no one-forms that have the same parity under $A,B,C$ and
$D$. Furthermore, the only two-forms with a fixed parity all have
negative parity and are spanned by $\bigl\{e^{16}\,,  e^{24}\,,
e^{35}\bigr\}$. The odd three-forms are spanned by $
\bigl\{e^{456}\,, e^{236}\,, e^{134}\,, e^{125}\bigr\}$.
Since $J$ and $\Omega_R$ must be odd under orientifold involutions
preserving the SU(3)-structure, we find that they must be of the
form
\subeq{\label{JOmexp}
\begin{align}
& J = ae^{16}+ be^{24} + ce^{35}\,,\\
&\Omega_R =v_1e^{456} + v_2e^{236} + v_3e^{134} + v_4e^{125}\,,
\end{align}}
with $a, b, c, v_1,\ldots, v_4$ real coefficients. With some abuse
of language we name $a, b, c$ the K\"ahler moduli and $v_1,\ldots,
v_4$ the complex structure moduli. Note that this implies
the calibration conditions
\begin{equation}
j_6\wedge \Omega_R=0=j_6\wedge J\, .
\end{equation}
The orientifold involutions also restrict the possible metric flux,
which must be even, or equivalently the possible group manifolds. In particular, the Lie
algebra should be of the following form
\eq{\spl{
& \d e^1 =f^1{}_{23}e^{23}+f^1{}_{45}e^{45}\,,\qquad \d e^2 =f^2{}_{13}e^{13}+f^2{}_{56}e^{56}\,,\\
& \d e^3 =f^3{}_{12}e^{12}+f^3{}_{46}e^{46}\,,\qquad \d e^4 =f^4{}_{36}e^{36}+f^4{}_{15}e^{15}\,,\label{algebra}\\
& \d e^5 =f^5{}_{14}e^{14}+f^5{}_{26}e^{26}\,,\qquad \d e^6
=f^6{}_{34}e^{34}+f^6{}_{25}e^{25}\,.
}}
As a consistency check one finds that $\d J$ indeed gives only rise
to odd three-forms and that $\d \Omega_R=0$ automatically. Furthermore,
the algebra is unipotent ($f^a{}_{ab}=0$) automatically. Unipotence is
a necessary condition for having a compact group manifold
(after the quotient by a discrete subgroup
$\Gamma$ if need be). The Jacobi identities, which are equivalent to the nilpotence $\d^2 e^i=0$,
impose further quadratic constraints on the $f$'s.

From
\begin{equation}
J\wedge J\wedge J = -6 \,abc \, e^{123456}\,,
\end{equation}
we find that (for our choice of orientation) $abc<0$ rendering all or
one of the coefficients $a, b, c$ negative.
In order to be able to properly normalise $I^2=-\mathbbm{1}$ with real $c$
in \eqref{complexstructure} we need furthermore
$v_1v_2v_3v_4>0$.
From equation (\ref{metricfromI}) we obtain the metric, which turns
out to be diagonal, consistent with even parity under the orientifold involutions
\begin{equation}
g=\frac{1}{\sqrt{v_1v_2v_3v_4}}\Bigl(a v_3v_4\,,\, -bv_2v_4\,,\,
cv_2v_3\,,\, -bv_1v_3\,,\, cv_1v_4\,,\, av_1v_2\Bigr)\,.
\end{equation}
With the metric available we can compute $\Omega_I=\star\Omega_R$
\begin{equation}
\Omega_I=\sqrt{v_1v_2v_3v_4}
\Bigl(v_1^{-1}\,e^{123}+v_2^{-1}\,e^{145}-v_3^{-1}\,e^{256}-v_4^{-1}e^{346}\Bigl)\,.
\end{equation}
The normalisation condition \eqref{normcond} leads to
\begin{equation}
\sqrt{v_1v_2v_3v_4}=-abc\,.
\end{equation}

The required parity under the orientifold involutions \eqref{oriinv}
will automatically imply that $W_4 = W_5 = 0$ and $W_1,W_2$ real so
that we indeed obtain a half-flat SU(3)-structure as advertised.
Furthermore, we can construct the remaining torsion classes from the
identities\footnote{To understand how the torsion classes depend on
all moduli is not too hard for these simple examples, but formulae
for more general cases have been derived in \cite{Ihl:2007ah}.}
\subeq{\al{ W_1&=-\tfrac{1}{6}\star_6 (\d J \wedge\Omega_I)\,,\\
W_2&=-\star \d\Omega_I+2W_1J\,,\\ W_3&=\d
J-\tfrac{3}{2}W_1\Omega_R\,.}}
The parity properties further imply the following relations
\eq{\spl{
&\d W_3=0\,, \qquad \qquad \d W_2 \wedge W_3 = 0 \,, \\
& W_2 \wedge W_3 = 0 \, , \qquad W_2 \wedge \star_6 W_3 = 0 \,.
}}

\subsection{de Sitter no-go theorems}

The above choice of orbifold/orientifold group is not unique and we
chose it for simplicity. But, interestingly, all other choices of
orbifold groups have shown not to admit any de Sitter solutions
according to \cite{Flauger:2008ad}.

The explicit form of the de Sitter no-go theorem of
\cite{Flauger:2008ad}, in our setup, implies that the matrix
\begin{equation}
F = \begin{pmatrix}
f^1{}_{45} &  f^1{}_{23} &-f^6{}_{34} & -f^6{}_{25} \\
f^2{}_{56} & -f^4{}_{36} & f^2{}_{13} & f^4{}_{15} \\
-f^3{}_{46}&  f^5{}_{26} & f^5{}_{14} & f^3{}_{12}
\end{pmatrix}
\end{equation}
should have at least two columns non-zero and three rows non-zero.

It is not straightforward to classify all six-dimensional
Lie-algebras that can be written as (\ref{algebra}). But once an
algebra is found it is easy to check the no-go theorem.
We postpone a full classification of groups and cosets that violate
the no-go theorem to the future. Sofar we have found four algebras
which fulfill the conditions
\begin{equation}
\SO(4)\,,\qquad \SO(3,1)\,,\qquad \SO(2,2)\,,\qquad s 1.2\,,
\end{equation}
where $s 1.2$ is the solvable algebra that appears in e.g. the
classification of \cite{Grana:2006kf}. Note that
$\SO(4)=\SU(2)\times\SU(2)$ and $\SO(2,2)=\SU(1,1)\times\SU(1,1)$.

The first and the last correspond to the algebras used for the de
Sitter solutions constructed in \cite{Caviezel:2008tf,
Flauger:2008ad}. The last three algebras correspond to non-compact
groups and one has to demonstrate that they can be compactified. We
do not discuss this here, but mention that this can depend on the
explicit solution, since the symmetries of the metric depend on the
expectation values for the metric scalars $a, b, c, v_1,\ldots, v_4$.

The algebras are explicitly given by\footnote{The basis for $\SO(4)$
is related to the basis used in e.g.~\cite{Koerber:2008rx} (which we
label $\tilde{e}^i$) as follows: $e^1 = \tilde{e}^1 +
\tilde{e}^4,e^2 = \tilde{e}^2 + \tilde{e}^5,e^3 = \tilde{e}^3 +
\tilde{e}^6,e^4 = \tilde{e}^2 - \tilde{e}^5, e^5 = \tilde{e}^3 -
\tilde{e}^6$ and $e^6 = \tilde{e}^1 - \tilde{e}^4$.}  \eq{\spl{
F_{\SO(4)} &= \frac{1}{2}\begin{pmatrix}
+1 & +1 & +1 & -1 \\
+1 & -1 & -1 & -1 \\
+1 & -1 & +1 & +1
\end{pmatrix}\,,\qquad
F_{\SO(2,2)} = \frac{1}{2}\begin{pmatrix}
+1 & +1 & +1 & -1 \\
-1 & +1 & +1 & +1 \\
+1 & -1 & +1 & +1
\end{pmatrix}\,,\\
F_{\SO(3,1)} &= \frac{1}{2}\begin{pmatrix}
-1 & +1 &+1 & -1 \\
-1 & -1 &-1 & -1 \\
-1&  -1 &+1 & +1
\end{pmatrix}\,,\qquad
 F_{s1.2} = \frac{1}{2}\begin{pmatrix}
-1 &  +1 & 0 & 0 \\
-1 & -1  & 0 & 0 \\
+1 &  +1 & 0 & 0
\end{pmatrix}\,,
}}
up to re-scalings, permutations, and other transformations in
$\GL(6,\Real)$ that preserve the form (\ref{algebra}). We expect
more algebras to exist that evade the no-go theorem of
\cite{Flauger:2008ad} and more details will be presented elsewhere
\cite{progress}.

\subsection{Constraints on the torsion classes}
\label{conscoeff}

With some explicit algebras at hand we can check the various
conditions for allowing the universal de Sitter solution of the
previous section. Let us recapitulate the conditions on the torsion classes
\subeq{\al{
W_2&=0\,,\label{constraint1}\\
\d \star_6 \hat{W}_3 & = c_1 J \wedge J  \, ,\label{constraint2} \\
(\hat{W}_3^2)_{ij}^+&=0\label{constraint3}\, , \\
 Q_1(\hat{W}_3,\hat{W}_3)&=c_2 \, Q_2(\hat{W}_3,\hat{W}_3)\,,\label{constraint4}\\
 Q_1(\hat{W}_3,\hat{W}_3)&=c_3 (\hat{W}_3)_{2,1}\,.\label{constraint5}
}}
Taking into account the restricted set of odd three-forms under the orientifold
involutions (see the discussion before \eqref{JOmexp}) we can verify
that constraint \eqref{constraint3} is automatic as well as \eqref{constraint4}
for $c_2=1$.

Likewise, constraint (\ref{constraint5}) can be simplified using the O-plane
ansatz. Let us write
\begin{equation}
W_3=\alpha_1 e^{456} + \alpha_2 e^{236} + \alpha_3e^{134} +
\alpha_4e^{125}\,,
\end{equation}
where we note that the $\alpha_i$ are not entirely arbitrary as $W_3$
needs to be a simple (2,1)+(1,2)-form.
After some algebra one finds that (\ref{constraint5}) leads to the
following four simple equations
\begin{equation}
2\left(\frac{3\alpha_i^2}{v_i^2}-\sum_{j\neq
i}\frac{\alpha_j^2}{v_j^2}\right)=c_3 w_3\frac{\alpha_i}{v_i}\,.
\end{equation}
One verifies that these equations are only consistent when
\begin{equation}
\label{qval}
c_3=\pm\frac{8}{\sqrt{3}} \quad \text{or} \quad c_3 = 0 \,.
\end{equation}
We do not consider the case $c_3=0$ as we have not found any
interesting results with that value. Furthermore, we can always take
the positive value for $c_3$ by allowing $w_3$ and $\hat{W}_3$ to
flip sign (under which the physical $W_3$ does not change).

The remaining constraint equations need to be imposed and fix the
geometric moduli $a, b, c, v_i$. Furthermore we found from the
analysis of section \ref{solana} that we need \eq{ \label{w3limits}
4.553 < w_3/W_1 < 3 \sqrt{3} \, . } We have investigated these
constraints for the four algebras listed above and only for
$\SU(2)\times\SU(2)$ we found that the constraints lead to a
solution. In all other cases one finds a complex solution or a
solution with a metric that is not positive-definite anymore. For
SO(3,1) we came close, but then \eqref{w3limits} was violated.

This implies that the de Sitter solution of \cite{Flauger:2008ad},
which was found for the s1.2 algebra, is not captured with our
ansatz. We have also checked that it can not be found from the more
extended ansatz  with $W_2\neq 0$. It would be interesting to
understand how many extra non-universal forms one needs to describe
this solution from a ten-dimensional viewpoint.

\section{The solution on $\SU(2)\times \SU(2)$}\label{SU2XSU2}

Let us now demonstrate that on the group manifold
$\SU(2)\times\SU(2)$ one can satisfy all the constraints
(\ref{constraint1}-\ref{constraint5},\ref{qval},\ref{w3limits}) above and construct a dS
solution.
The geometric moduli values for which $\SU(2)\times\SU(2)$ obeys the
constraints (\ref{constraint1}-\ref{constraint5}) are given by
\begin{equation}
a=-b=c\,,\qquad v_1=v_2=v_4\,,\qquad v_3=\frac{a^6}{(v_1)^3}\, .
\end{equation}
For these values we have explicitly
\subeq{\al{
g & = \text{diag}\left(\tfrac{a^4}{(v_1)^2},\tfrac{(v_1)^2}{a^2},\tfrac{a^4}{(v_1)^2},\tfrac{a^4}{(v_1)^2},\tfrac{(v_1)^2}{a^2},\tfrac{(v_1)^2}{a^2}\right) \, , \\
W_1 & = \frac{a^6 + v_1^4}{4 \, a^5 v_1} \, , \qquad W_2 = 0 \, , \\
W_3 & = \frac{a^6-3 (v_1)^4}{8 \,a^5 (v_1)^4} \left[ (v_1)^4
(e^{456}+e^{236}+e^{125})-3 \,a^6 e^{134}\right]  \, , \qquad w_3 =
-\tfrac{\sqrt{3}(a^6-3 (v_1)^4)}{4 a^5 v_1} \, , }} where we have
chosen the sign of $w_3$ such that $c_3=8/\sqrt{3}$.\footnote{That
this configuration satisfies the constraints
(\ref{constraint1}-\ref{constraint3}) and \ref{constraint5} can be
easily seen as follows. Consider the exchanges \eq{ p_1: e^1 \leftrightarrow
e^4, e^6 \leftrightarrow e^2 \qquad p_2: e^1 \leftrightarrow e^3,
e^6 \leftrightarrow e^5 \, . } One easily sees that the only even
two-form under these interchanges is $J$, while from the
three-forms odd under the orientifold involutions, the ones also odd
under this symmetry group are spanned by $\Omega_R$ and $W_3$.
Furthermore, the structure constants and thus the exterior
derivative are also odd. The above constraints now follow from the
transformation properties under these exchanges.}

We find furthermore that
\eq{
w_3/W_1=\frac{\sqrt{3}(3 (v_1)^4
-a^6)}{a^6+(v_1)^4}
}
takes values in the interval $-\sqrt{3}<w_3/W_1 < 3\sqrt{3}$ so that we can satisfy
\eqref{w3limits} and find dS solutions (as well as a Minkowski solution at $w_3/W_1=4.553$
and AdS solutions below that value).  Note that since the endpoint $w_3/W_1=3 \sqrt{3}$ would correspond to
$(v_1)^4/a^6 \rightarrow \infty$ we cannot actually construct the
Minkowski solution (\ref{Minkowski}). It is amusing to see the geometry of SU(2)$\times$SU(2) seems
to know about the type IIA supergravity equations of motion in that
the upper bound for $w_3/W_1$ on
SU(2)$\times$SU(2) corresponds to the upper bound for any type of solutions in figure \ref{plotw3}.

\begin{figure}
\centering
\psfrag{M2}{$\scriptstyle M^2/(f_1)^2$}
\psfrag{f2}{$\scriptstyle f_2/f_1$}
\includegraphics[scale=0.5]{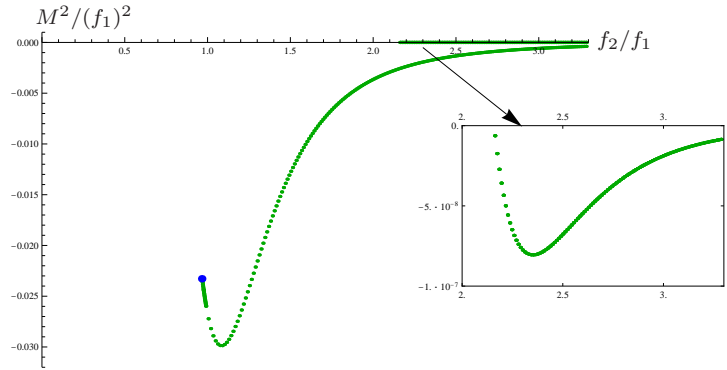}
\caption{Negative eigenvalues of the mass matrix $M^2/(f_1)^2$ for
our line of dS solutions.}
\label{plotspectrum}
\end{figure}
Using $\mathcal{N}=1$ supergravity techniques along the lines of \cite{Caviezel:2008tf}
it is possible to calculate the scalar potential and obtain the eigenvalues of
the $(14\times 14)$ mass matrix for the left-invariant fluctuations around our dS solutions.
We found that there are always unstable directions. In figure \ref{plotspectrum}
we make a plot of the negative eigenvalues. In
particular we have one tachyonic direction for $f_2/f_1< 2.16$ and three tachyonic directions
beyond this values, as shown in the figure 3. The two extra
tachyons that appear beyond $f_2/f_1= 2.16$ have the same
negative (mass)$^2$ and since it is small compared to the other
masses, we have zoomed in on it in the figure.

\section{Discussion}\label{discussion}

We have established the fact that the class of universal de Sitter
solutions exists by at least providing one explicit example: an
unstable de Sitter solution on an orientifold of $\SU(2)\times \SU(2)$.
This solution turns out to be the ten-dimensional lift of the
solution found in \cite{Caviezel:2008tf}. Furthermore, the other
explicit solution, coming from (an orientifold) of the solvmanifold
s1.2 \cite{Flauger:2008ad} is not universal, which demonstrates that
the class of universal de Sitter solutions does not cover all of the
classical de Sitter solutions. But, as we argued in the
introduction, we believe that most natural solutions are of this
form and we hope to report on more examples in a future work
\cite{progress}.

Obviously, the understanding of classical de Sitter solutions is
very incomplete, and sofar no conclusion can be drawn regarding the
existence of phenomenologically viable solutions. The requirements
for phenomenological viability are plenty, and some requirements can
be dropped for other purposes. Especially if we want a simple de
Sitter solution for the sake of understanding holography, or more
general, quantum gravity in de Sitter space-time, we can drop the
requirement for a small cosmological constant, a decoupling of KK
modes, etc. The most important requirement is perturbative
stability. The two examples coming from $\SU(2)\times\SU(2)$ and the
solvmanifold s1.2 are both perturbative unstable. This is not
surprising since the lack of supersymmetry does not protect one from
tachyons in the spectrum. Since most simple models have order 10
moduli, one has to be rather lucky that all of them have positive
mass. In case one is allowed to think that, in the absence of susy,
there is an equal chance for a field direction to be unstable or
stable, then one is forced to conclude that stable solutions must
exist if there are enough classical de Sitter solutions in the
landscape.  But, the existence of meta-stable non-susy AdS solutions
seems to demonstrate that one should not apply a ``50-50'' reasoning
for all field directions to be stable.

When it comes to stability with respect to the light degrees of
freedom it is useful to consult the results that were obtained
directly in four-dimensional supergravity theories, without the
concern of a higher-dimensional origin. Investigations on this
mainly focus on extended gauged supergravities \cite{
Kallosh:2001gr, deRoo:2002jf, deRoo:2003rm, Fre:2002pd,
Ogetbil:2008tk, GomezReino:2008bi, Roest:2009tt, Dibitetto:2010rg},
and some on $\mathcal{N}=1$ supergravity \cite{Covi:2008ea,
Covi:2008cn, deCarlos:2009qm, deCarlos:2009fq}. These results
indicate that in theories with $\mathcal{N}>2$ metastable solutions
do not exist, while for theories with $\mathcal{N}\leq 2$ metastable
solutions exist, but are not generic. For those metastable examples
for which the ten-dimensional origin seems understood the
solutions are non-geometric \cite{deCarlos:2009qm}, which is
problematic since the supergravity limit might be invalid. The known
geometric examples \cite{Caviezel:2008tf, Flauger:2008ad} are
solutions of four-dimensional $\mathcal{N}=1$ supergravity. Although
the solutions are unstable, they do evade the known no-go theorems
for stable de Sitter solutions in $\mathcal{N}=1$ supergravity
\cite{Covi:2008ea, Covi:2008cn}.

Finally a word on directions that deserve further investigation.
Obviously, it would be interesting to find more explicit simple de
Sitter solutions, in order to be able to understand how generic
unstable solutions are and whether meta-stable solutions can be
found at all. With the universal ansatz there is no reason to
restrict to group and coset manifolds and we can also look at
non-homogeneous manifolds. One could also be interested in simple de
Sitter solutions in higher dimensions than four. In higher
dimensions we expect less moduli and this simplifies the situation,
and, for pure theoretical purposes de Sitter solutions in any
dimension are of interest. Another way to obtain more classical de
Sitter solutions would be by looking at IIB supergravity. Recently
it has been shown in reference \cite{Caviezel:2009tu} that moduli
stabilisation at tree-level in IIB AdS vacua is possible as well,
and more importantly, an unstable dS critical point of the effective
potential was found. This opens up the interesting possibility of
finding tree-level dS vacua in IIB.

Perhaps a more pressing problem than stability is the issue of
the backreaction of the sources. Orientifolds are really localised objects and
smearing them leads to lower-dimensional supergravity effective
actions, but the consistency of smearing from a string theory point
of view has not been shown \cite{Douglas:2010rt}. This issue
arises for some of the AdS solutions as well, and it is important to sort it out.

\section*{Acknowledgements}
We benefitted from useful discussions and ongoing collaboration with
Shajid Haque, Gary Shiu and Timm Wrase.  We also thank Timm Wrase
for comments on the manuscript.  T.V.R likes to thank Tine De Smedt
for some help in editing the pictures.

U.D.\ is supported by the
Swedish Research Council (VR) and the G\"{o}ran Gustafsson
Foundation. P.K.\ is a Postdoctoral Fellow of the FWO -- Vlaanderen.
The work of P.K.\ is further supported in part by the FWO --
Vlaanderen project G.0235.05 and in part by the Federal Office for
Scientific, Technical and Cultural Affairs through the
`Interuniversity Attraction Poles Programme Belgian Science Policy'
P6/11-P.  T.V.R.\ is supported by the G\"{o}ran Gustafsson
Foundation.

\bibliography{universal}

\providecommand{\href}[2]{#2}\begingroup\raggedright\begin{thebibliography}{10}

\bibitem{DeWolfe:2005uu}
O.~DeWolfe, A.~Giryavets, S.~Kachru and W.~Taylor,  {\em {Type IIA moduli
  stabilization}}, JHEP {\bf 07} (2005) 066
[\href{http://www.arXiv.org/abs/hep-th/0505160}{{\tt hep-th/0505160}}].

\bibitem{Maldacena:2000mw}
J.~M. Maldacena and C.~N\'u\~{n}ez,  {\em {Supergravity description of field
  theories on curved manifolds and a no go theorem}}, Int. J. Mod. Phys. {\bf
  A16} (2001) 822--855
[\href{http://www.arXiv.org/abs/hep-th/0007018}{{\tt hep-th/0007018}}].

\bibitem{Hertzberg:2007wc}
M.~P. Hertzberg, S.~Kachru, W.~Taylor and M.~Tegmark,  {\em {Inflationary
  constraints on type IIA string theory}}, JHEP {\bf 12} (2007) 095
[\href{http://www.arXiv.org/abs/0711.2512}{{\tt 0711.2512}}].

\bibitem{Silverstein:2007ac}
E.~Silverstein,  {\em {Simple de Sitter solutions}}, Phys. Rev. {\bf D77}
  (2008) 106006
[\href{http://www.arXiv.org/abs/0712.1196}{{\tt 0712.1196}}].

\bibitem{Haque:2008jz}
S.~S. Haque, G.~Shiu, B.~Underwood and T.~Van~Riet,  {\em {Minimal simple de
  Sitter solutions}}, Phys. Rev. {\bf D79} (2009) 086005
[\href{http://www.arXiv.org/abs/0810.5328}{{\tt 0810.5328}}].

\bibitem{Caviezel:2008tf}
C.~Caviezel, P.~Koerber, S.~K\"ors, D.~L\"ust, T.~Wrase and M.~Zagermann,  {\em
  {On the cosmology of type IIA compactifications on SU(3)-structure
  manifolds}}, JHEP {\bf 04} (2009) 010
[\href{http://www.arXiv.org/abs/0812.3551}{{\tt 0812.3551}}].

\bibitem{Flauger:2008ad}
R.~Flauger, S.~Paban, D.~Robbins and T.~Wrase,  {\em {Searching for slow-roll
  moduli inflation in massive type IIA supergravity with metric fluxes}}, Phys.
  Rev. {\bf D79} (2009) 086011
[\href{http://www.arXiv.org/abs/0812.3886}{{\tt 0812.3886}}].

\bibitem{Danielsson:2009ff}
U.~H. Danielsson, S.~S. Haque, G.~Shiu and T.~Van~Riet,  {\em {Towards
  classical de Sitter solutions in string theory}}, JHEP {\bf 09} (2009) 114
[\href{http://www.arXiv.org/abs/0907.2041}{{\tt 0907.2041}}].

\bibitem{Wrase:2010ew}
T.~Wrase and M.~Zagermann,  {\em {On classical de Sitter Vacua in string
  theory}},
\href{http://www.arXiv.org/abs/1003.0029}{{\tt 1003.0029}}.

\bibitem{Covi:2008cn}
L.~Covi, M.~G\'omez-Reino, C.~Gross, J.~Louis, G.~A. Palma and C.~A. Scrucca,
  {\em {Constraints on modular inflation in supergravity and string theory}},
  JHEP {\bf 08} (2008) 055
[\href{http://www.arXiv.org/abs/0805.3290}{{\tt 0805.3290}}].

\bibitem{Lust:2004ig}
D.~L\"ust and D.~Tsimpis,  {\em {Supersymmetric AdS$_4$ compactifications of
  IIA supergravity}}, JHEP {\bf 02} (2005) 027
[\href{http://www.arXiv.org/abs/hep-th/0412250}{{\tt hep-th/0412250}}].

\bibitem{tomasiellocosets}
A.~Tomasiello,  {\em {New string vacua from twistor spaces}}, Phys. Rev. {\bf
  D78} (2008) 046007
[\href{http://www.arXiv.org/abs/0712.1396}{{\tt 0712.1396}}].

\bibitem{Koerber:2008rx}
P.~Koerber, D.~L\"ust and D.~Tsimpis,  {\em {Type IIA AdS$_4$ compactifications
  on cosets, interpolations and domain walls}}, JHEP {\bf 07} (2008) 017
[\href{http://www.arXiv.org/abs/0804.0614}{{\tt 0804.0614}}].

\bibitem{Caviezel:2008ik}
C.~Caviezel, P.~Koerber, S.~K\"{o}rs, D.~Tsimpis, D.~L\"{u}st and M.~Zagermann,
   {\em {The effective theory of type IIA AdS$_4$ compactifications on
  nilmanifolds and cosets}}, Class. Quant. Grav. {\bf 26} (2009) 025014
[\href{http://www.arXiv.org/abs/0806.3458}{{\tt 0806.3458}}].

\bibitem{Cassani:2009ck}
D.~Cassani and A.-K. Kashani-Poor,  {\em {Exploiting $\mathcal{N}$=2 in
  consistent coset reductions of type IIA}}, Nucl. Phys. {\bf B817} (2009)
  25--57
[\href{http://www.arXiv.org/abs/0901.4251}{{\tt 0901.4251}}].

\bibitem{cveticnk1}
K.~Behrndt and M.~Cveti\u{c},  {\em {General $\mathcal{N}$ = 1 supersymmetric
  flux vacua of (massive) type IIA string theory}}, Phys. Rev. Lett. {\bf 95}
  (2005) 021601
[\href{http://www.arXiv.org/abs/hep-th/0403049}{{\tt hep-th/0403049}}].

\bibitem{progress}
U.~H. Danielsson, S.~S. Haque, P.~Koerber, G.~Shiu, T.~Van~Riet and T.~Wrase.
\newblock {Work in progress}.

\bibitem{bedulli-2007-4}
L.~Bedulli and L.~Vezzoni,  {\em The Ricci tensor of SU(3)-manifolds}, J. Geom.
  Phys. {\bf 4} (2007) 1125
[\href{http://www.arXiv.org/abs/math/0606786}{{\tt math/0606786}}].

\bibitem{Ali:2006gd}
T.~Ali and G.~B. Cleaver,  {\em {The Ricci curvature of half-flat manifolds}},
  JHEP {\bf 05} (2007) 009
[\href{http://www.arXiv.org/abs/hep-th/0612171}{{\tt hep-th/0612171}}].

\bibitem{Lust:2008zd}
D.~L\"ust, F.~Marchesano, L.~Martucci and D.~Tsimpis,  {\em {Generalized
  non-supersymmetric flux vacua}}, JHEP {\bf 11} (2008) 021
[\href{http://www.arXiv.org/abs/0807.4540}{{\tt 0807.4540}}].

\bibitem{dimitriosextrasol}
D.~L\"ust and D.~Tsimpis,  {\em {Classes of AdS$_4$ type IIA/IIB
  compactifications with SU(3)$\times$SU(3) structure}}, JHEP {\bf 04} (2009)
  111
[\href{http://www.arXiv.org/abs/0901.4474}{{\tt 0901.4474}}].

\bibitem{Koerber:2010rn}
P.~Koerber and S.~K\"ors,  {\em {A landscape of non-supersymmetric AdS vacua on
  coset manifolds}},
\href{http://www.arXiv.org/abs/1001.0003}{{\tt 1001.0003}}.

\bibitem{Li:2009pf}
W.~Li, T.~Nishioka and T.~Takayanagi,  {\em {Some no-go theorems for string
  duals of non-relativistic Lifshitz-like theories}}, JHEP {\bf 10} (2009) 015
[\href{http://www.arXiv.org/abs/0908.0363}{{\tt 0908.0363}}].

\bibitem{Blaback:2010pp}
J.~Bl{\aa}b\"ack, U.~H. Danielsson and T.~Van~Riet,  {\em {Lifshitz backgrounds
  from 10d supergravity}}, JHEP {\bf 02} (2010) 095
[\href{http://www.arXiv.org/abs/1001.4945}{{\tt 1001.4945}}].

\bibitem{chiossal}
S.~Chiossi and S.~Salamon,  {\em The intrinsic torsion of SU(3) and G$_2$
  structures}, Ann. Mat. Pura e Appl. {\bf 282} (1980) 35--58
[\href{http://www.arXiv.org/abs/math/0202282}{{\tt math/0202282}}].

\bibitem{Koerber:2007hd}
P.~Koerber and D.~Tsimpis,  {\em {Supersymmetric sources, integrability and
  generalized-structure compactifications}}, JHEP {\bf 08} (2007) 082
[\href{http://www.arXiv.org/abs/0706.1244}{{\tt 0706.1244}}].

\bibitem{Bergshoeff:2001pv}
E.~Bergshoeff, R.~Kallosh, T.~Ort\'{\i}n, D.~Roest and A.~Van~Proeyen,  {\em
  {New formulations of $D=10$ supersymmetry and D8-O8 domain walls}}, Class.
  Quant. Grav. {\bf 18} (2001) 3359--3382
[\href{http://www.arXiv.org/abs/hep-th/0103233}{{\tt hep-th/0103233}}].

\bibitem{deCarlos:2009qm}
B.~de~Carlos, A.~Guarino and J.~M. Moreno,  {\em {Complete classification of
  Minkowski vacua in generalised flux models}},
\href{http://www.arXiv.org/abs/0911.2876}{{\tt 0911.2876}}.

\bibitem{Grana:2006kf}
M.~Gra\~na, R.~Minasian, M.~Petrini and A.~Tomasiello,  {\em {A scan for new
  $\mathcal{N}=1$ vacua on twisted tori}}, JHEP {\bf 05} (2007) 031
[\href{http://www.arXiv.org/abs/hep-th/0609124}{{\tt hep-th/0609124}}].

\bibitem{Ihl:2007ah}
M.~Ihl, D.~Robbins and T.~Wrase,  {\em {Toroidal orientifolds in IIA with
  general NS-NS Fluxes}}, JHEP {\bf 08} (2007) 043
[\href{http://www.arXiv.org/abs/0705.3410}{{\tt 0705.3410}}].

\bibitem{Kallosh:2001gr}
R.~Kallosh, A.~D. Linde, S.~Prokushkin and M.~Shmakova,  {\em {Gauged
  supergravities, de Sitter space and cosmology}}, Phys. Rev. {\bf D65} (2002)
  105016
[\href{http://www.arXiv.org/abs/hep-th/0110089}{{\tt hep-th/0110089}}].

\bibitem{deRoo:2002jf}
M.~de~Roo, D.~B. Westra and S.~Panda,  {\em {de Sitter solutions in $N = 4$
  matter coupled supergravity}}, JHEP {\bf 02} (2003) 003
[\href{http://www.arXiv.org/abs/hep-th/0212216}{{\tt hep-th/0212216}}].

\bibitem{deRoo:2003rm}
M.~de~Roo, D.~B. Westra, S.~Panda and M.~Trigiante,  {\em {Potential and
  mass-matrix in gauged $N = 4$ supergravity}}, JHEP {\bf 11} (2003) 022
[\href{http://www.arXiv.org/abs/hep-th/0310187}{{\tt hep-th/0310187}}].

\bibitem{Fre:2002pd}
P.~Fr\'e, M.~Trigiante and A.~Van~Proeyen,  {\em {Stable de Sitter vacua from
  $\mathcal{N} = 2$ supergravity}}, Class. Quant. Grav. {\bf 19} (2002)
  4167--4194
[\href{http://www.arXiv.org/abs/hep-th/0205119}{{\tt hep-th/0205119}}].

\bibitem{Ogetbil:2008tk}
O.~\"Ogetbil,  {\em {Stable de Sitter vacua in 4-dimensional supergravity
  originating from 5 dimensions}}, Phys. Rev. {\bf D78} (2008) 105001
[\href{http://www.arXiv.org/abs/0809.0544}{{\tt 0809.0544}}].

\bibitem{GomezReino:2008bi}
M.~G\'omez-Reino, J.~Louis and C.~A. Scrucca,  {\em {No metastable de Sitter
  vacua in $\mathcal{N}=2$ supergravity with only hypermultiplets}}, JHEP {\bf
  02} (2009) 003
[\href{http://www.arXiv.org/abs/0812.0884}{{\tt 0812.0884}}].

\bibitem{Roest:2009tt}
D.~Roest and J.~Rosseel,  {\em {de Sitter in extended supergravity}}, Phys.
  Lett. {\bf B685} (2010) 201--207
[\href{http://www.arXiv.org/abs/0912.4440}{{\tt 0912.4440}}].

\bibitem{Dibitetto:2010rg}
G.~Dibitetto, R.~Linares and D.~Roest,  {\em {Flux compactifications, gauge
  algebras and de Sitter}},
\href{http://www.arXiv.org/abs/1001.3982}{{\tt 1001.3982}}.

\bibitem{Covi:2008ea}
L.~Covi, M.~G\'omez-Reino, C.~Gross, J.~Louis, G.~A. Palma and C.~A. Scrucca,
  {\em {de Sitter vacua in no-scale supergravities and Calabi-Yau string
  models}}, JHEP {\bf 06} (2008) 057
[\href{http://www.arXiv.org/abs/0804.1073}{{\tt 0804.1073}}].

\bibitem{deCarlos:2009fq}
B.~de~Carlos, A.~Guarino and J.~M. Moreno,  {\em {Flux moduli stabilisation,
  supergravity algebras and no-go theorems}}, JHEP {\bf 01} (2010) 012
[\href{http://www.arXiv.org/abs/0907.5580}{{\tt 0907.5580}}].

\bibitem{Caviezel:2009tu}
C.~Caviezel, T.~Wrase and M.~Zagermann,  {\em {Moduli stabilization and
  cosmology of type IIB on SU(2)- structure orientifolds}},
\href{http://www.arXiv.org/abs/0912.3287}{{\tt 0912.3287}}.

\bibitem{Douglas:2010rt}
M.~R. Douglas and R.~Kallosh,  {\em {Compactification on negatively curved
  manifolds}},
\href{http://www.arXiv.org/abs/1001.4008}{{\tt 1001.4008}}.

\end{thebibliography}\endgroup
\bibliographystyle{utphysmodb}
\end{document}